\documentstyle{article}
\begin{document}

\noindent{\large{\bf Atomic number fluctuations in a falling cold atom cloud}}

\vspace{0.5cm}

{\sc Astrid Lambrecht, Elisabeth Giacobino and Serge Reynaud} \\

{\it Laboratoire Kastler Brossel, UPMC, ENS, CNRS}

{\it Universit\'e Pierre et Marie Curie, case 74, F-75252 Paris, France}\\

({\sc Quantum Semiclassical Optics} {\bf 8}, 457 (1996))\\

\begin{abstract}
We evaluate the effective number of atoms in experiments where a probe laser
beam with a Gaussian transverse profile passes through an atomic medium
consisting in a cold atom cloud released from a magneto-optical trap.
Considering the case where the initial distribution is a Gaussian function
of position and of velocity, we give a quantitative description of the
temporal variation of the effective number while the cloud is exploding and
falling down. We discuss the two cases where the effective number is defined
from the linear and nonlinear phaseshifts respectively. We also evaluate the
fluctuations of the effective atomic number by calculating their correlation
functions and the associated noise spectra. We finally estimate the effect
of these fluctuations on experiments where the probe beam passes through a
cavity containing the atomic cloud.
\end{abstract}

\section{\bf{Introduction}}

In recent years, the development of magneto-optical traps \cite{MOT} has
opened the way to the use of cold atoms for various purposes such as atomic
interferometry, and nonlinear or quantum optics.\ In many of these
experiments, a cloud of cold atoms is first trapped and then released when
the laser beams and the magnetic field of the trap are turned off.\ The
experiment is thus performed while the cloud is undergoing a ballistic
expansion due to the velocity distribution in the trap and falling down
under the influence of gravity.\ Models using a Monte-Carlo simulation of
the evolution of the atomic positions and velocities have been developed to
analyze the free fall of the cloud of cold atoms.\ They have been used to
obtain informations about the initial position and velocity distribution in
the trap from fluorescence measurements, particularly to evaluate
temperature in the trap when it is released and to investigate deviations
from a Maxwellian velocity distribution (\cite{Drewsen} and references
therein).

Here, we want to concentrate on the category of experiments where the
falling cloud of cold atoms is used in spectroscopy, nonlinear or quantum
optics experiments.\ In this case, the cloud interacts with a probe laser
beam that passes through it.\ A crucial point for these experiments lies in
the definition of an effective number of atoms contributing to the
interaction with the probe beam and also to variation of this effective
atomic number as a function of time.

A first aim of the present paper is to provide a definition of the effective
number of atoms in the probe beam, as well as a precise modelisation of its
temporal evolution.\ We will show that this is a quite involved problem in
the general case, and in particular that different definitions may be given,
depending upon the specific physical effect chosen as a reference for the
evaluation of the atomic number.\ We will discuss the example where the
linear phase produced by the atomic medium is chosen as a reference, as well
as the case where the nonlinear phase shift is considered.\ This general
discussion will lead to simpler conclusions in the two limiting cases where
either the transverse size of the probe beam is much smaller than the cloud
size or the Rayleigh length characteristic of the divergence of the probe
beam is much larger than the cloud size.

A further problem is that this effective number of atoms has fluctuations.\
These fluctuations must be added to the fluctuations usually evaluated with
the number of atoms kept constant and they may have an influence in
spectroscopy, metrology or quantum optics experiments.\ In this class of
experiments indeed, the signal is often measured with such a high precision
that any excess noise produced by the atomic medium can appear as a limiting
factor for the sensitivity.

This is why in a second part of this paper we will evaluate atomic number
fluctuations by calculating their characteristic correlation functions.\ We
will show in particular that it is possible to calculate time dependent
noise spectra, when the transverse size of the probe beam is much smaller
than the cloud size.\ In this case indeed, the correlation time, expected to
be of the order of the time of flight through the probe beam, is much
smaller than the typical variation time of the mean number of atoms, of the
order of the time of flight through the cloud, so that the effective atomic
number may be considered as a quasistationary random variable.

The theoretical developments of the present paper are in particular
essential for the understanding and interpretation of experimental results
we obtained recently \cite{Squeezing94}.\ The experiment consists in the
measurement of quantum fluctuations of a weak probe beam passing through a
cloud of cold atoms expanding and falling down after it has been released
from a standard magneto-optical trap.\ To enhance the interaction between
atoms and light, the atoms are placed inside an optical cavity.\ The
variation of the number of atoms leads to a continuous variation of the
linear phaseshift of the laser beam, which is sufficient to scan the cavity
across resonance.\ Due to the non-stationary character of the measurement of
quantum fluctuations, the knowledge of the time variation of the number of
atoms plays a crucial role in the interpretation of experimental results.\
The fit with the experimental results, using the theory developed in the
present paper, will be presented in a further publication \cite{Squeezing95}%
.\ While a variation of the mean atomic number scans the cavity detuning
through a variation of the linear phase shift, atomic number fluctuations
result in fluctuations of the detuning which have an effect on the
fluctuations of the probe beam which are measured in the experiment.\ If
they are large enough, they may enter exponentially in the expression for
the field fluctuations after interaction, and therefore produce excess noise
not only at the frequencies contained in their own noise spectrum, but also
at higher frequencies through a multiplicative noise processing \cite{MNoise}%
.\ The evaluation of atomic number fluctuations contained in the present
paper are particularly useful to delineate under which conditions these
fluctuations degrade squeezing measurements.

As already pointed out by Weiss et al. \cite{Weiss}, the density
distribution of a freely falling atom cloud can be calculated analytically
at any instant, if a Gaussian profile in the trap is assumed for position
density as well as for velocity distribution.\ Experiments by Drewsen et al. 
\cite{Drewsen} have shown this assumption to be justified if multiple
scattering effects become negligeable, that means for large clouds and atoms
with a temperature in the 100 $\mu $K to mK range \cite{Ketterle}.\ These
conditions are met in the experiment that we are modelizing \cite
{Squeezing94}.\ When the number of trapped atoms is deduced from their
fluorescence signal, the detectivity function does not have a Gaussian
profile and does in general not lead to analytical results.\ In the case
studied in this paper, the experimental determination of the number of atoms
relies on a phase shift measurement of a probe laser beam.\ It may thus be
expected that the detectivity function is given by the Gaussian profile of
the laser beam, leading to analytical solutions for the mean atomic number
and for correlation functions of atomic number fluctuations.

In Section 2 of this paper, we describe our model in more detail and show
how the effective number of atoms is deduced from phase shift measurements,
accounting for the transverse mode structure of the probe beam.\ From this,
we analyze in Section 3 the expression of the case of a linear phase shift
measurement, considering the general case of an atomic motion under
gravity.\ In Section 4, we address the problem of a saturated phase shift.\
The correlation functions and the noise spectrum of atomic number
fluctuations are then calculated in Section 5.\ When evaluating
fluctuations, we restrict our attention to the particular case of the atomic
number deduced from the linear phase shift since we are mainly interested in
potential effects of linear phase fluctuations.\ We finally give an estimate
of the effect of the latter on the probe-cavity detuning in Section 6.

\section{\bf{Description of the model}}

As already discussed, we consider that the number of atoms is monitored
through a phase measurement performed on a probe laser beam.\ Hence, this
number will be obtained as an integral of the atomic density over the
volume, with a weight function fitting the Gaussian profile of the laser
beam.

To determine precisely this weight function, we introduce the Gaussian mode $%
u\left( {\bf r}\right) $ (${\bf r}$ stands for the three-dimensional
position ($x,y,z$)) describing the intracavity field propagating along the $%
x $ direction and coming onto the atoms:

\begin{equation}
u\left( {\bf r}\right) =\sqrt{\frac 2\pi }\frac 1{w\left( x\right) }\exp
\left[ -\frac{y^2+z^2}{w^2\left( x\right) }-i\varphi \left( {\bf r}%
\right) \right]  \label{1}
\end{equation}
$w\left( x\right) $ is the position-dependent beam size:

\begin{equation}
w^2\left( x\right) =w_0^2\left( 1+\frac{x^2}{l_R^2}\right)  \label{2}
\end{equation}
\begin{equation}
l_R=\frac{\pi w_0^2}\lambda  \label{2b}
\end{equation}
where $w_0$ is the beam waist, $l_R$ the Rayleigh divergence length and $%
\lambda $ the laser wavelength. The field phase $\varphi \left( {\bf r}%
\right) $ is the sum of contributions representing propagation phase and
curvature of the wave at longitudinal position $x$:

\begin{equation}
\varphi \left( {\bf r}\right) =-\frac{2\pi x}\lambda +\arctan%
\left( \frac x{l_R}\right) -\frac \pi \lambda \frac{\left( y^2+z^2\right) x}{%
x^2+l_R^2}  \label{3}
\end{equation}
The squared modulus of the Gaussian mode:

\begin{equation}
\left| u\left( {\bf r}\right) \right| ^2=\frac{f\left( {\bf r}\right) 
}{S\left( x\right) }  \label{4}
\end{equation}
may be written in terms of a Gaussian weight function $f\left( {\bf r}%
\right) $ normalized to unity on the beam axis and an effective beam section 
$S$:

\begin{equation}
f\left( {\bf r}\right) =\exp \left[ \frac{-2\left( y^2+z^2\right) }{%
w^2\left( x\right) }\right]  \label{5}
\end{equation}

\begin{equation}
S\left( x\right) =\frac{\pi w^2\left( x\right) }2  \label{6}
\end{equation}
The normalisation of the Gaussian mode $u\left( {\bf r}\right) $ has been
chosen so that:

\begin{equation}
\int \left| u\left( {\bf r}\right) \right| ^2dy\ dz=1  \label{7}
\end{equation}
This relation holds for any position $x$ along the propagation axis.

We will write the incident field seen at time $t$ by the atoms located at
point ${\bf r}$ in the interaction representation with respect to the
laser frequency $\omega _L$:

\begin{equation}
{\em E}\left( {\bf r,}t\right) =e^{-i\omega _Lt}E\left( {\bf r,}%
t\right) +e^{i\omega _Lt}E\left( {\bf r,}t\right) ^{\dagger }  \label{8}
\end{equation}
where $E\left( {\bf r,}t\right) $ is the product of the Gaussian profile $%
u\left( {\bf r}\right) $ by a mode amplitude $A\left( t\right) $: 
\begin{equation}
E\left( {\bf r,}t\right) =\sqrt{\frac{\hbar \omega _L}{2\varepsilon _0c}}%
A\left( t\right) u\left( {\bf r}\right)  \label{8b}
\end{equation}
This definition is such that $A\left( t\right) ^{\dagger }A\left( t\right) $
is the number of photons passing through a beam section per unit time.

As explained below, the field radiated by the atoms will have a different
transverse profile as a consequence of the nonlinear character of
light-matter interaction. The Gaussian mode $u\left( {\bf r}\right) $ is
the lowest order mode $u_{00}$ of the family of orthogonal Hermite modes.\
We will consider that the higher-order Hermite modes $u_{mn}$ (with $m\neq 0$
or $n\neq 0$) are not resonant at the same frequency as the lowest order
mode $u_{00}$.\ Hence, any field radiated by the atoms into these
higher-order modes will not be efficiently coupled back onto the atoms and
can therefore be disregarded.\ We will then calculate the field amplitude
radiated into the fundamental mode by a mere projection.

The effect of an optically thin atomic layer of length $dx$ may be described
by a local modification $dE\left( {\bf r}\right) $of the field:

\begin{equation}
E\left( {\bf r}\right) \rightarrow E\left( {\bf r}\right) +dE\left( 
{\bf r}\right)  \label{9}
\end{equation}
with \cite{Heidmann85}:

\begin{equation}
dE\left( {\bf r}\right) =-\frac{3\lambda ^2}{4\pi }\rho \left( {\bf r}%
\right) \alpha \left( {\bf r}\right) E\left( {\bf r}\right) dx
\label{e10}
\end{equation}
Note that we have treated the motion of atoms as quasistatic and neglected
any consequence of the Doppler effect.\ As a result, we have used the atomic
density $\rho \left( {\bf r}\right) $ integrated over the velocity
distribution. $\alpha \left( {\bf r}\right) $ is the atomic
polarizability measured as a dimensionless number:

\begin{equation}
\alpha \left( {\bf r}\right) =\frac{\alpha _l}{1+2s\left( {\bf r}%
\right) }  \label{11}
\end{equation}
The linear polarizability $\alpha _l$ is a function of the dimensionless
detuning $\delta $ between laser frequency $\omega _L$ and atomic resonance
frequency $\omega _0$ normalized to the decay rate $\gamma $ of the atomic
dipole:

\begin{equation}
\alpha _l=\frac 1{1+i\delta }  \label{12}
\end{equation}

\begin{equation}
\delta =\frac{\omega _0-\omega _L}\gamma  \label{13}
\end{equation}
The saturation parameter $s$ is proportional to the local laser intensity $%
\left| \beta \left( {\bf r}\right) \right| ^2$ measured as a
dimensionless number:

\begin{equation}
s\left( {\bf r}\right) =\frac{\left| \beta \left( {\bf r}\right)
\right| ^2}{1+\delta ^2}  \label{e14}
\end{equation}
\begin{equation}
\beta \left( {\bf r}\right) =\frac{d_0E\left( {\bf r}\right) }{\hbar
\gamma }=\sqrt{\frac{3\lambda ^2}{4\pi }}u\left( {\bf r}\right) \frac A{%
\sqrt{\gamma }}  \label{14b}
\end{equation}
where $d_0$ is the matrix element of the atomic dipole.

As a first step, we consider that the saturation of polarizability may be
neglected; we will come back to the nonlinear case later on.\ The number of
atoms is then evaluated by monitoring the linear phase shift produced by the
atomic cloud, with $dE\left( {\bf r}\right) $ written as:

\begin{equation}
dE\left( {\bf r}\right) =-\frac{3\lambda ^2}{4\pi }\rho \left( {\bf r}%
\right) \alpha _lE\left( {\bf r}\right) dx  \label{15}
\end{equation}
Even in this linear case, the radiated field $dE\left( {\bf r}\right) $
does not have the same spatial variation as the incident field $E\left( 
{\bf r}\right) $ when the atomic density $\rho \left( {\bf r}\right) $
is position-dependent.\ Ignoring the field radiated into higher-order modes,
one gets the modification of the field amplitude $dA$ in the Gaussian mode
by projecting $dE\left( {\bf r}\right) $ onto $u\left( {\bf r}\right) $%
:

\begin{equation}
\sqrt{\frac{\hbar \omega _L}{2\varepsilon _0c}}dA=\int dE\left( {\bf r}%
\right) \ u^{*}\left( {\bf r}\right) \ dy\ dz  \label{e16}
\end{equation}
Assuming that the modification of the field by the whole atomic cloud
remains small (assumption of an optically thin atomic cloud), we obtain this
modification by summing up the contributions of all the atomic layers as:

\begin{equation}
\frac{dA}A=-\frac{3\lambda ^2}{4\pi }\alpha _l\ \sigma  \label{e17}
\end{equation}
with :

\begin{equation}
\sigma =\int \left| u\left( {\bf r}\right) \right| ^2\ \rho \left( 
{\bf r}\right) \ d{\bf r}  \label{18}
\end{equation}
The modification of the field intensity is:

\begin{equation}
d\left( A^{\dagger }A\right) =A^{\dagger }dA+\left( dA^{\dagger }\right)
A=-A^{\dagger }A\frac{3\lambda ^2}{2\pi }\frac \sigma {1+\delta ^2}
\label{19}
\end{equation}
As the absorption cross section is $\left( 3\lambda ^2/2\pi \right) $ at
resonance and is divided by $\left( 1+\delta ^2\right) $ otherwise, this
equation shows that $\sigma $ must be interpreted as the effective number of
atoms present in the detection beam per unit beam section.\ This
interpretation may be made more precise by rewriting $\sigma $ as:

\begin{equation}
\sigma =\int \frac{dn\left( x\right) }{S\left( x\right) }  \label{e20}
\end{equation}
where $dn\left( x\right) $ is the number of atoms present in the detection
beam in a thin layer of length $dx$, that is the number obtained by
integrating the atomic density $\rho \left( {\bf r}\right) $ over the
beam area with a Gaussian weight function $f\left( {\bf r}\right) $:

\begin{equation}
\frac{dn\left( x\right) }{dx}=\int f\left( {\bf r}\right) \rho \left( 
{\bf r}\right) dy\ dz  \label{e21}
\end{equation}
This effective number has to be distinguished from the number of atoms
present in the whole cloud in a thin layer of length $dx$, which is obtained
by integrating the atomic density $\rho \left( {\bf r}\right) $:

\begin{equation}
\frac{d{\cal N}\left( x\right) }{dx}=\int \rho \left( {\bf r}\right)
dy\ dz  \label{e22}
\end{equation}

The atomic medium considered in this paper consists in a freely falling
cloud of cold atoms.\ We will assume that the phase space distribution $\pi
\left( {\bf r,v,}0\right) $, at time $t=0$ chosen as the beginning of the
free fall, is a Gaussian function of position as well as of velocity:

\begin{equation}
\pi \left( {\bf r,v,}0\right) =\frac{{\cal N}}{\left( 2\pi \sigma
_r\sigma _v\right) ^3}\exp \left( -\frac{{\bf r}^2}{2\sigma _r^2}-\frac{%
{\bf v}^2}{2\sigma _v^2}\right)  \label{23}
\end{equation}
${\cal N}$ is the total number of atoms while $\sigma _{\rm r}$
measures the radius of the initial cloud and $\sigma _v$ the thermal
velocity related to the trap temperature $T$ through:

\begin{equation}
\sigma _v^2=\frac{k_BT}m  \label{e24}
\end{equation}
As already mentioned in the introduction, this Gaussian approximation gives
a rather good description of the trap distribution \cite{Drewsen,Ketterle}.\
It presents the great advantage that it will lead to analytical expressions
for the integrals encountered in the evaluation of the mean atomic number
and of atomic number fluctuations, at least in the case where the Rayleigh
divergence length is larger than the cloud size.\ For the sake of
simplicity, the initial cloud has been considered to be centrered on the
detection beam, and the detection beam supposed to propagate in horizontal
direction.\ Note however that releasing these two simplifying assumptions
would not change the Gaussian character of the atomic density, which is the
key point of the following calculations.

The motion of atoms at later times t is determined by the laws of free fall,
which preserve the phase space volume:

\begin{equation}
d{\bf r}\left( t\right) d{\bf v}\left( t\right) =d{\bf r}\left(
0\right) d{\bf v}\left( 0\right)  \label{25}
\end{equation}
The time-evolution of the phase space distribution $\pi \left( {\bf r,v,}%
t\right) $ is deduced from the conservation of the number of atoms
(Liouville's theorem):

\begin{equation}
\pi \left( {\bf r+v}t+\frac 12{\bf g}t^2,{\bf v}+{\bf g}%
t,t\right) =\pi \left( {\bf r,v,}0\right)   \label{26}
\end{equation}
with ${\bf g}$ oriented downwards in the $z$ direction, or equivalently:

\begin{equation}
\pi \left( {\bf r,v,}t\right) =\pi \left( {\bf r-v}t+\frac 12{\bf g}%
t^2,{\bf v-g}t,0\right)   \label{27}
\end{equation}
The atomic density is defined as the integral over velocity of the phase
space distribution:

\begin{equation}
\rho \left( {\bf r,}t\right) =\int \pi \left( {\bf r,v,}t\right) d%
{\bf v}  \label{28}
\end{equation}
This Gaussian integral is easily evaluated \cite{Weiss}:

\begin{equation}
\rho \left( {\bf r,}t\right) =\frac{{\cal N}}{\left[ 2\pi \left(
\sigma _r^2+\sigma _v^2t^2\right) \right] ^{3/2}}\exp \left[ -\frac{\left( 
{\bf r-}\frac 12{\bf g}t^2\right) ^2}{2\left( \sigma _r^2+\sigma
_v^2t^2\right) }\right]   \label{29}
\end{equation}
For example, the atomic density at the initial center of the cloud $\left( 
{\bf r}={\bf 0}\right) $ is:

\begin{equation}
\rho \left( {\bf 0},t\right) =\frac{{\cal N}}{\left[ 2\pi \sigma
_r^2\right] ^{3/2}}\left[ \frac{\tau _r^2}{\tau _r^2+t^2}\right] ^{3/2}\exp
\left[ \frac{-t^4}{\tau _g^2\left( \tau _r^2+t^2\right) }\right]  \label{e30}
\end{equation}
with:

\begin{equation}
\tau _r=\frac{\sigma _r}{\sigma _v}  \label{e31}
\end{equation}

\begin{equation}
\tau _g=\frac{2\sqrt{2}\sigma _v}{\left| {\bf g}\right| }  \label{e32}
\end{equation}
This shows that the atomic cloud undergoes a ballistic expansion on a time
scale $\tau _r$, the time of flight through the trap radius $\sigma _r$ for
an atom flying at mean thermal velocity $\sigma _v$, while at the same time
the cloud is falling down under the influence of gravity on a time scale $%
\tau _g$.\ When the time scale $\tau _g$ is longer than $\tau _r$, the
effect of gravity may be disregarded for short times ($t\ll \tau _g$) and
the atomic density varies mainly as a consequence of ballistic expansion.\
Gravity leads to an exponential decrease at longer times ($t\gg \tau _g$)
and the variation of the atomic density may be rewritten in this case:

\begin{equation}
\rho \left( {\bf 0},t\right) \approx \frac{{\cal N}}{\left[ 2\pi
\sigma _r^2\right] ^{3/2}}\left[ \frac{\tau _r^2}{\tau _r^2+t^2}\right]
^{3/2}\exp\left[ -\frac{t^2}{\tau _g^2}\right]  \label{33}
\end{equation}
A third time scale $\tau _w$ will be found in the following, which will
measure the time of flight through the detection beam:

\begin{equation}
\tau _w\left( x\right) =\frac{w\left( x\right) }{2\sigma _v}  \label{34}
\end{equation}

\section{\bf{Effective number of atoms deduced from the linear phase
shift}}

Using the results of the previous section, the linear phase shift of the
field due to the interaction with the atomic cloud may be written in terms
of equations (\ref{e17}, \ref{e20}, \ref{e21}) where we have introduced a
time-dependence of the effective atomic number per beam section $\sigma
\left( t\right) $ in order to account for the quasistatic time-variation of
the atomic density $\rho \left( {\bf r},t\right) $:

\begin{equation}
\frac{dA}A=-\alpha _l\frac{3\lambda ^2}{4\pi }\sigma \left( t\right)
\label{35}
\end{equation}

\begin{equation}
\sigma \left( t\right) =\int \frac{dn\left( x,t\right) }{S\left( x\right) }
\label{e36}
\end{equation}

\begin{equation}
\frac{dn\left( x,t\right) }{dx}=\int f\left( {\bf r}\right) \rho \left( 
{\bf r,}t\right) dy\ dz  \label{e37}
\end{equation}
The latter integral over the transverse variables $y$ and $z$ has a Gaussian
character and can be readily performed.\ The resulting expression for the
number of atoms $dn(x,t)$ present at time $t$ in the detection beam in a
thin layer of length $dx$ may be written:

\begin{equation}
dn\left( x,t\right) =\frac{d{\cal N}\left( x,t\right) w^2\left( x\right) 
}{4\left( \sigma _r^2+\sigma _v^2t^2\right) +w^2\left( x\right) }\exp \left[
-\frac{\frac 12g^2t^4}{4\left( \sigma _r^2+\sigma _v^2t^2\right) +w^2\left(
x\right) }\right]  \label{e38}
\end{equation}
$d{\cal N}\left( x,t\right) $ is the number of atoms present at time $t$
in the whole cloud in a thin layer of length $dx$ (cf. definition \ref{e22}):

\begin{equation}
\frac{d{\cal N}\left( x,t\right) }{dx}=\frac{{\cal N}}{\sqrt{2\pi
\left( \sigma _r^2+\sigma _v^2t^2\right) }}\exp \left[ -\frac{x^2}{2\left(
\sigma _r^2+\sigma _v^2t^2\right) }\right]  \label{39}
\end{equation}
The integral over the longitudinal variable $x$ keeps a Gaussian character
either when the beam size is much smaller than the trap radius ($w\ll \sigma
_r$) or when the Rayleigh divergence length is much larger than the cloud
size ($\sigma _r\ll l_R$).\ We discuss these two limiting cases in the next
paragraphs, and recall that both assumptions are fulfilled in the
experimental situation of reference \cite{Squeezing94}.

We consider first the limiting case $w\ll \sigma _r$, but with $\sigma
_r/l_R $ arbitrary, where the number of atoms $dn(x,t)$ given by equation (%
\ref{e38}) is proportional to $w^2\left( x\right) $, and therefore also to
the beam section $S(x)$.\ The $x$-dependent factor $S(x)$ disappears in the
expression (\ref{e36}) of $\sigma \left( t\right) $ which is thus a Gaussian
integral.\ Its evaluation leads to:

\begin{equation}
\sigma \left( t\right) =\frac{{\cal N}}{2\pi \sigma _v^2\left( \tau
_r^2+t^2\right) }\exp \left[ -\frac{t^4}{\tau _g^2\left( \tau
_r^2+t^2\right) }\right]  \label{e40}
\end{equation}
where $\tau _r$ and $\tau _g$ are the already introduced time scales which
correspond respectively to free flight through the trap and free fall
(compare with equation \ref{e30}).

In the case where the Rayleigh divergence length is much larger than the
cloud size ($\sigma _r\ll l_R$ with $w/\sigma _r$ arbitrary), the $x$%
-dependence of the beam section $S$ can be ignored and $\sigma \left(
t\right) $ is again a Gaussian integral. For an arbitrary beam size $w$, the
time of flight through the probe beam $\tau _w$ becomes a relevant
parameter.\ After the evaluation of the integral, we obtain $\sigma \left(
t\right) $ as:

\begin{equation}
\sigma \left( t\right) =\frac{{\cal N}}{2\pi \sigma _v^2\left( \tau
_r^2+\tau _w^2+t^2\right) }\exp \left[ -\frac{t^4}{\tau _g^2\left( \tau
_r^2+\tau _w^2+t^2\right) }\right]  \label{e41}
\end{equation}
where $\tau _w$ is the time of flight through the probe beam.\ When both
assumptions, a large Rayleigh divergence length and a small beam size ($w\ll
\sigma _r$ and $\sigma _r\ll l_R$), are valid, expression (\ref{e41}) is
identical to expression (\ref{e40}) which is valid for any ratio $\sigma
_r/l_R$.

As already discussed with the help of equation (\ref{e30}), the effect of
gravity may be disregarded for times $t\ll \tau _g$ when $\tau _r\ll \tau _g$%
.\ It follows that $n(t)$ thus varies as a Lorentzian function of time, as a
consequence of ballistic explosion.\ Gravity leads to an exponential
decrease at longer times.\ In the particular case $\tau _w\ll \tau _r\ll
\tau _g$ for example, these various regimes of time-dependence of $\sigma
\left( t\right) $ may be described in a single formula:

\begin{equation}
\sigma \left( t\right) =\frac{{\cal N}}{2\pi \sigma _v^2\left( \tau
_r^2+t^2\right) }\exp \left[ -\frac{t^2}{\tau _g^2}\right]   \label{42}
\end{equation}
This condition $\tau _r\ll \tau _g$ corresponds to a high temperature limit,
since it may be rewritten from equations (\ref{e24}\ref{e31},\ref{e32}):

\begin{equation}
k_BT\gg \frac{m\sigma _r\left| {\bf g}\right| }{2\sqrt{2}}
\end{equation}
In the opposite low temperature limit, the variation of $\sigma \left(
t\right) $ is mainly determined for all times by the exponential factor
which describes the free fall of the cloud under the effect of gravity.

\section{\bf{Effective number of atoms deduced from the nonlinear phase
shift}}

In the previous section, we have studied the field modification in a linear
regime, and deduced an effective number of atoms from a linear phase shift.\
We now come to the general case where the atomic polarizability may be
saturated.

The transverse variation of $dE\left( {\bf r}\right) $ is then completely
different from the transverse variation of the incident field $E\left( 
{\bf r}\right) $ as a consequence of the dependence of the saturation
parameter (see equations \ref{e10}-\ref{e14}):

\begin{equation}
dE\left( {\bf r}\right) =-\frac{\alpha _l}{1+2s\left( {\bf r}\right) }%
\frac{3\lambda ^2}{4\pi }\rho \left( {\bf r}\right) E\left( {\bf r}%
\right) dx  \label{43}
\end{equation}

\begin{equation}
s\left( {\bf r}\right) =s_m\left( x\right) f\left( {\bf r}\right)
\label{44}
\end{equation}
$s_m$ is the saturation parameter evaluated on the beam axis:

\begin{equation}
s_m\left( x\right) =\frac{3\lambda ^2}{4\pi S\left( x\right) }\frac{\left|
A\right| ^2}\gamma \frac 1{1+\delta ^2}  \label{45}
\end{equation}
As in the linear case, we ignore the field radiated into higher-order modes
and we obtain the modification of the field amplitude $dA$ projected onto
the Gaussian mode through equation (\ref{e16}):

\begin{equation}
\frac{dA}A=-\alpha _l\frac{3\lambda ^2}{4\pi }\sigma _s\left( t\right)
\label{46}
\end{equation}
where $\sigma _s$ is now modified by saturation:

\begin{equation}
\sigma _s\left( t\right) =\int \frac{dn_s\left( x,t\right) }{S\left(
x\right) }  \label{47}
\end{equation}

\begin{equation}
\frac{dn_s\left( x,t\right) }{dx}=\int \frac{f\left( {\bf r}\right) }{%
1+2s_m\left( x\right) f\left( {\bf r}\right) }\rho \left( {\bf r,}%
t\right) dy\ dz  \label{48}
\end{equation}

In order to evaluate $\sigma _s$, we will first expand $dn_s$ in powers of
the saturation parameter:

\begin{equation}
dn_s\left( x,t\right) =\sum _{k=0}^{\infty}\left(
-2s_m\left( x\right) \right) ^kdn^{\left( 1+k\right) }\left( x,t\right)
\label{49}
\end{equation}
where:

\begin{equation}
\frac{dn^{\left( j\right) }\left( x,t\right) }{dx}=\int f\left( {\bf r}%
\right) ^j\rho \left( {\bf r,}t\right) dy\ dz  \label{50}
\end{equation}
$dn^{\left( 1\right) }(x,t)$ is exactly the expression (\ref{e37}) of the
effective number of atoms $dn(x,t)$ in the detection beam in a thin layer of
length $dx$, as it was evaluated in the previous section.\ Due to the
Gaussian shape of function $f({\bf r})$, function $f({\bf r})^j$ has
the same expression as $f({\bf r)}$ with a modified value of the beam
size parameter, precisely with $w^2\left( x\right) $ replaced by $\left(
w^2\left( x\right) /j\right) $. Therefore, $dn^{\left( j\right) }(x,t)$
corresponds to the expression (\ref{e38}) of $dn(x,t)$ evaluated for a
modified beam size:

\begin{equation}
dn^{\left( j\right) }\left( x,t\right) =\frac{d{\cal N}\left( x,t\right) 
\frac{w^2\left( x\right) }j}{4\left( \sigma _r^2+\sigma _v^2t^2\right) +%
\frac{w^2\left( x\right) }j}\exp \left[ -\frac{\frac 12g^2t^4}{4\left(
\sigma _r^2+\sigma _v^2t^2\right) +\frac{w^2\left( x\right) }j}\right]
\label{51}
\end{equation}

These expressions are greatly simplified in the limiting case of a small
beam size ($w\ll \sigma _r$), where $dn^{\left( j\right) }(x,t)$ is
proportional to $w^2\left( x\right) $, so that:

\begin{equation}
dn^{\left( j\right) }\left( x,t\right) =\frac{dn\left( x,t\right) }j
\label{52}
\end{equation}
Assuming furthermore that the Rayleigh divergence length is large ($\sigma
_r\ll l_R$), one may disregard the $x$-dependence of the on-axis saturation
parameter $s_m$ and obtain a closed analytical expression for $dn_s\left(
x,t\right) $ and therefore for $\sigma _s$:

\begin{equation}
\sigma _s\left( t\right) =\frac{\sigma \left( t\right) }{2s_m}\ln \left(
1+2s_m\right)  \label{e53}
\end{equation}
This expression may then be used to derive the bistability relation which
connects the mean field sent into the cavity and the mean field inside the
cavity (see for example \cite{Drummond81}).\ An important consequence of
this equation is that $\sigma _s\left( t\right) $ and $\sigma \left(
t\right) $ have exactly the same time dependence.\ It follows that the
evaluations of the number of atoms obtained from the linear and non linear
phase shift agree as soon as the transverse structure of the mode has been
properly accounted for.\ It is however worth to stress that this property
relies on the two simplifying assumptions $w\ll \sigma _r$ and $\sigma _r\ll
l_R$ which have been used to derive equation (\ref{e53}), and does no longer
hold in a more general situation.

\section{\bf{Fluctuations of the number of atoms}}

Up to now, we have studied only the mean value of the field modification in
the atomic medium.\ We now want to evaluate the fluctuations of this
modification which are associated with fluctuations of the number of atoms
coupled to the field. As shown by the previous discussions, the definition
of such a number is not simple in the general case.\ This is why we will
study the fluctuations of the number of atoms with the help of simplifying
assumptions.

First, we will restrict our attention to the regime of a linear
polarisability where the field modification is related to the quantity $%
\sigma \left( t\right) $ given by equation (\ref{e20}).\ We are indeed
interested mainly in fluctuations of the linear phase shift, which is
usually much larger than the nonlinear one. We will then consider that the
Rayleigh divergence length is much larger than the cloud size ($\sigma _r\ll
l_R$), so that the beam area $S$ is independent of $x$ and equation (\ref
{e20}) may be rewritten in terms of an effective number $n(t)$ of atoms in
the probe beam at time $t$:

\begin{equation}
\sigma \left( t\right) =\frac{n\left( t\right) }S  \label{54}
\end{equation}
with:

\begin{equation}
n\left( t\right) =\int dn\left( x,t\right) =\int f\left( {\bf r}\right)
\rho \left( {\bf r},t\right) d{\bf r}  \label{55}
\end{equation}
It would be possible to study the fluctuations of $\sigma \left( t\right) $
in the general case, using the same techniques as presented below.\ However,
it will be more instructive to discuss the fluctuations of $n\left( t\right) 
$, which is a dimensionless number and is expected to have a nearly
Poissonian statistics.\ For completeness, note that in the limiting case
studied in the present section ($\sigma _r\ll l_R$), $n\left( t\right) $ is
simply the product of $\sigma \left( t\right) $ by the beam section $S$.

We come now to the discussion of atomic number fluctuations.\ To this aim,
we notice that the atomic density $\rho \left( {\bf r},t\right) $ may be
considered as an average value taken over the random distribution of the
atomic positions, precisely over the random values of the initial positions
and velocities:

\begin{equation}
\rho \left( {\bf r},t\right) =\left\langle P\left( {\bf r},t\right)
\right\rangle  \label{56}
\end{equation}

\begin{equation}
P\left( {\bf r},t\right) =\sum _{i=1}^{\cal N} 
\delta \left( {\bf r-r}_i\left( t\right) \right) =
\sum_{i=1}^{\cal N}\delta \left( {\bf r-r}_i\left( 0\right) -
{\bf v}_i\left( 0\right) t-\frac 12{\bf g}t^2\right)   \label{57}
\end{equation}
where $i$ labels the atoms ($i=1\ldots {\cal N}$ with a total number of
atoms ${\cal N}$ in the cloud); ${\bf r}_i\left( t\right) $ is the
position of the $i$th atom at time $t$; ${\bf r}_i\left( 0\right) $ and $%
{\bf v}_i\left( 0\right) $ are its initial position and velocity.\
Therefore, $n\left( t\right) $ may be considered as the mean value of a
random number $N\left( t\right) $:

\begin{equation}
N\left( t\right) =\sum _{i=1}{\cal N} f\left( 
{\bf r}_i\left( t\right) \right)  \label{58}
\end{equation}
This random number has fluctuations which, as already discussed in the
introduction, may reveal themselves as fluctuations of the linear phase
shift and then of the cavity-detuning parameter.

In order to discuss the statistical properties of these numbers, we will
consider that the atoms are statistically uncorrelated in the initial
distribution.\ We will define covariances of random variables according to
the general prescription:

\begin{equation}
\left\langle u,v\right\rangle =\left\langle uv\right\rangle -\left\langle
u\right\rangle \left\langle v\right\rangle  \label{59}
\end{equation}
$d{\cal N}_0$ is the random variable associated with the number of atoms
in an elementary volume $d{\bf r}$ $d{\bf v}$ of the phase space at
time $t=0$. We consider it to be a Poisson random variable:

\begin{equation}
\left\langle d{\cal N}_0,d{\cal N}_0\right\rangle =\left\langle d%
{\cal N}_0\right\rangle =\pi \left( {\bf r,v},0\right) d{\bf r}\ d%
{\bf v}  \label{60}
\end{equation}
while different $d{\cal N}_0$ and $d{\cal N}_0^{\prime }$ evaluated
for non overlapping elementary volumes are uncorrelated random variables:

\begin{equation}
\left\langle d{\cal N}_0,d{\cal N}_0^{\prime }\right\rangle =0
\label{61}
\end{equation}

Before coming to the discussion of the fluctuations of the effective atomic
number in the detection beam, we want to recall briefly how these
assumptions are commonly used to derive Poisson statistics.\ Usually, one
studies a number of atoms integrated in a finite volume:

\begin{equation}
N_0=\int d{\cal N}_0f_v\left( {\bf r}\right)  \label{62}
\end{equation}
where the function $f_v$ is either l inside the detection volume or 0
otherwise; this function therefore obeys the property $f_v^2=f_v$.\ By
summing up the variances corresponding to the various elementary volumes,
one deduces that the integrated number $N_0$ has a Poissonian variance:

\begin{equation}
\left\langle N_0,N_0\right\rangle =\int \left\langle d{\cal N}%
_0\right\rangle f_v^2\left( {\bf r}\right) =\left\langle N_0\right\rangle
\label{63}
\end{equation}
The same conclusion is reached for the number of atoms integrated in a
finite volume after a free fall during a time $t$:

\begin{equation}
N\left( t\right) =\int d{\cal N}_0\ f_v\left( {\bf r+v}t+\frac 12%
{\bf g}t^2\right)   \label{64}
\end{equation}

\begin{equation}
\left\langle N\left( t\right) ,N\left( t\right) \right\rangle =\int
\left\langle d{\cal N}_0\right\rangle \ f_v^2\left( {\bf r+v}t+\frac 12%
{\bf g}t^2\right) =\left\langle N\left( t\right) \right\rangle 
\label{65}
\end{equation}

The case studied in the present paper does not lead to such a simple
conclusion since the effective number is defined with a Gaussian weight
function which does not obey the property $f^2=f$.\ By summing up the
variances corresponding to the various elementary volumes, one obtains the
variance of this effective number to be:

\begin{equation}
\left\langle N\left( t\right) ,N\left( t\right) \right\rangle =\int
f^2\left( {\bf r+v}t+\frac 12{\bf g}t^2\right) \pi \left( {\bf r,v}%
,0\right) d{\bf r}\ d{\bf v}  \label{66}
\end{equation}
By changing variables from $\left\{ {\bf r}\left( 0\right) ,{\bf v}%
\left( 0\right) \right\} $ to $\left\{ {\bf r}\left( t\right) ,{\bf v}%
\left( t\right) \right\} $ and integrating over the velocity distribution,
this variance can be rewritten as:

\begin{equation}
\left\langle N\left( t\right) ,N\left( t\right) \right\rangle =\int
f^2\left( {\bf r}\right) \ \rho \left( {\bf r},t\right) \ d{\bf r}
\label{67}
\end{equation}
while the mean value is:

\begin{equation}
\left\langle N\left( t\right) \right\rangle =\int f\left( {\bf r}\right)
\ \rho \left( {\bf r},t\right) \ d{\bf r}  \label{68}
\end{equation}
The mean value is obtained as the product of the expression (\ref{e41})
evaluated previously for $\sigma \left( t\right) $ with the effective beam
section $\left( S=\pi w^2/2\right) $:

\begin{equation}
\left\langle N\left( t\right) \right\rangle =\frac{{\cal N}\tau _w^2}{%
\tau _r^2+\tau _w^2+t^2}\exp \left[ -\frac{t^4}{\tau _g^2\left( \tau
_r^2+\tau _w^2+t^2\right) }\right]  \label{69}
\end{equation}
The variance has the same expression evaluated for a modified beam size
parameter ($w^2$ replaced by $\left( w^2/2\right) $, i.e. $\tau _w^2$
replaced by $\left( \tau _w^2/2\right) $):

\begin{equation}
\left\langle N\left( t\right) ,N\left( t\right) \right\rangle =\frac{%
{\cal N}\tau _w^2}{2\tau _r^2+\tau _w^2+2t^2}\exp \left[ -\frac{2t^4}{%
\tau _g^2\left( 2\tau _r^2+\tau _w^2+2t^2\right) }\right]  \label{70}
\end{equation}
The variance is therefore smaller than the mean value, which implies that
the distribution of the random variable $N\left( t\right) $ has always a
sub-Poissonian character:

\begin{equation}
\left\langle N\left( t\right) ,N\left( t\right) \right\rangle <\left\langle
N\left( t\right) \right\rangle  \label{71}
\end{equation}
In the limiting case where the detection beam waist $w$ is much smaller than
the trap radius $\sigma _r$, the variance is simply half the mean value:

\begin{equation}
\left\langle N\left( t\right) ,N\left( t\right) \right\rangle =\frac
12\left\langle N\left( t\right) \right\rangle  \label{72}
\end{equation}

We now look for a dynamical characterization of temporal correlations rather
than for a static statistical description.\ Precisely, we wish to evaluate
the covariance function for the numbers of atoms $N(t)$ and $N(t\prime )$
present in the detection beam at two different times.\ By summing up the
variances corresponding to the various elementary volumes, one obtains this
covariance function as:

\begin{equation}
\left\langle N\left( t\right) ,N\left( t^{\prime }\right) \right\rangle
=\int f\left( {\bf r+v}t+\frac 12{\bf g}t^2\right) f\left( {\bf r+v}%
t^{\prime }+\frac 12{\bf g}t^{\prime 2}\right) \pi \left( {\bf r,v,}%
0\right) d{\bf r}\ d{\bf v}  \label{e73}
\end{equation}
Since the positions at times $t$ and $t^{\prime }$ are linear functions of
the initial position and initial velocity, this expression appears as a
six-dimensional Gaussian integral which may be readily evaluated.\ We will
write the covariance function:

\begin{equation}
C_{NN}\left( T,\tau \right) =\left\langle N\left( t\right) ,N\left(
t^{\prime }\right) \right\rangle   \label{74}
\end{equation}
as a function of the mean time $T$ and of the delay $\tau $ between the two
time parameters $t$ and $t^{\prime }$:

\begin{equation}
T=\frac{t+t^{\prime }}2\qquad \qquad \qquad \qquad \tau =t-t^{\prime }
\label{75}
\end{equation}
The time $T$ will be associated with the global time variation of the atomic
number during the free fall of the cloud while the delay $\tau $ will rather
correspond to the correlation between numbers evaluated at different times.

The explicit evaluation of the Gaussian integral (\ref{e73}) gives the
following expression:

\begin{equation}
C_{NN}\left( T,\tau \right) =n\left( 0\right) L\left( T,\tau \right) \exp
\left[ -M\left( T,\tau \right) \right]   \label{e76}
\end{equation}
where:

\begin{equation}
n\left( 0\right) =\frac{{\cal N}\tau _w^2}{\tau _r^2+\tau _w^2}
\label{77}
\end{equation}

\begin{equation}
L\left( T,\tau \right) =\frac{\tau _w^2\left( \tau _r^2+\tau _w^2\right) }{%
2\tau _w^2T^2+\left( \tau _r^2+\frac{\tau _w^2}2\right) \left( \tau ^2+2\tau
_w^2\right) }  \label{78}
\end{equation}

\begin{equation}
M\left( T,\tau \right) =\frac{\left( T^2+\frac{\tau ^2}4\right) ^2\left(
\tau ^2+2\tau _w^2\right) +4\left( \tau _r^2+\frac{\tau _w^2}2\right)
T^2\tau ^2}{\tau _g^2\left[ 2\tau _w^2T^2+\left( \tau _r^2+\frac{\tau _w^2}%
2\right) \left( \tau ^2+2\tau _w^2\right) \right] }  \label{e79}
\end{equation}
The expression of the variance of the number $N(T)$ evaluated at a given
time $T$ is recovered at $\tau =0$:

\begin{equation}
C_{NN}\left( T,0\right) =\left\langle N\left( T\right) ,N\left( T\right)
\right\rangle   \label{80}
\end{equation}

Instead evaluating the exact expressions (\ref{e76}-\ref{e79}), we will
rather concentrate on the limiting case where the detection waist is much
smaller than the trap radius. This assumption implies in particular that the
correlation time $\tau _w$ is much smaller than the global fall time $\tau
_r $.\ This will allow us to treat the number of atoms as a quasistationary
random variable.\ Considering only not too long delays $\tau \ll \tau _r$
and not too short fall times $T\gg \tau _w$, we may rewrite the covariance
function in terms of two scaled time parameters $T/\tau _r$ and $\tau /\tau
_w$:

\begin{equation}
C_{NN}\left( T,\tau \right) =n\left( 0\right) L\left( T,\tau \right) \exp
\left[ -\zeta \left( a_T-b_TL\left( T,\tau \right) \right) \right] 
\label{81}
\end{equation}
with:

\begin{equation}
n\left( 0\right) =\frac{{\cal N}\tau _w^2}{\tau _r^2}  \label{82}
\end{equation}

\begin{equation}
L\left( T,\tau \right) =\frac 1{\left( \frac \tau {\tau _w}\right) ^2+\alpha
_T^2}  \label{83}
\end{equation}
\begin{equation}
\alpha _T^2=2\left[ 1+\left( \frac T{\tau _r}\right) ^2\right]  \label{83b}
\end{equation}

\begin{equation}
a_T=\left( \frac T{\tau _r}\right) ^2\left[ 4+\left( \frac T{\tau _r}\right)
^2\right]  \label{84}
\end{equation}
\begin{equation}
b_T=2\left( \frac T{\tau _r}\right) ^2\left[ 2+\left( \frac T{\tau
_r}\right) ^2\right] ^2  \label{84b}
\end{equation}

\begin{equation}
\zeta =\frac{\tau _r^2}{\tau _g^2}  \label{85}
\end{equation}
In the limiting case $\zeta \ll 1$ where the time scale characteristic of
gravity is long, the covariance function is essentially a Lorentzian
function.\ More generally when including the effect of gravity, it may be
expanded in terms of powers of this Lorentzian function:

\begin{equation}
C_{NN}\left( T,\tau \right) =n\left( 0\right) e^{-\zeta a_T}
\sum_{k=0}^{\infty} \left( \zeta b_T\right) ^k\frac{L^{1+k}}{k!}
\label{e86}
\end{equation}

Since the correlation time, of the order of $\tau _w$, is much smaller than
the characteristic time of variation of the mean values, of the order of $%
\tau _r$, the number of detected atoms may be considered as a
quasistationary random variable and it is therefore possible to characterize
its fluctuations by a noise spectrum $S_{NN}\left( T,\omega \right) $:

\begin{equation}
S_{NN}\left( T,\omega \right) =\int d\tau ~~e^{i\omega \tau }C_{NN}\left(
T,\tau \right)   \label{87}
\end{equation}
This spectrum is obtained through a Fourier transformation with respect to
the delay $\tau $ and it slowly depends upon the time of fall $T$.\ It has
the same definition as the ambiguity function of radar theory \cite
{Ambiguity} which is the analog for time-frequency distributions of the
Wigner distribution for position-momentum distributions \cite{Wigner}.\
Because of the quasistationary character of the random variable $N\left(
t\right) $, it will remain positive for all values of the parameters, as the
noise spectrum for a stationary random variable.\ The spectrum may be
written as the product of the variance at time $T$ by a normalized spectrum:

\begin{equation}
S_{NN}\left( T,\omega \right) =C_{NN}\left( T,0\right) \bar{S}
\left( T,\omega \right) =\frac 12n\left( T\right) \bar{S}
_{NN}\left( T,\omega \right)   \label{e88}
\end{equation}

\begin{equation}
\int \frac{d\omega }{2\pi }\bar{S}_{NN}\left( T,\omega \right) =1
\label{89}
\end{equation}

In the limiting case $\zeta \ll 1$ , the effect of gravity upon the
correlation function $C_{NN}\left( T,\tau \right) $ may be disregarded, so
that the latter is a Lorentzian function of $\tau $ and the noise spectrum
has an exponential shape:

\begin{equation}
S_{NN}\left( T,\omega \right) =n\left( 0\right) \frac{\pi \tau _w}{\alpha _T}%
e^{-\alpha _T\omega \tau _w}  \label{90}
\end{equation}
The normalized spectrum:

\begin{equation}
\bar{S}_{NN}\left( T,\omega \right) =\pi \alpha _T\tau
_we^{-\alpha _T\omega \tau _w}  \label{91}
\end{equation}
has a linewidth $\left( \alpha _T\tau _w\right) ^{-1}$ and a peak value $%
\left( \pi \alpha _T\tau _w\right) $ mainly determined by the time of flight 
$\tau _w$ through the probe beam.

In the general case of an arbitrary $\zeta $, the noise spectrum may be
deduced from expression (\ref{e86}) of the correlation function (using
formulas 8.432.5 and 8.468 in ref. \cite{Gradshteyn}) as:

\begin{equation}
S_{NN}\left( T,\omega \right) =n\left( 0\right) \frac{\pi \tau _w}{\alpha _T}%
e^{-\alpha _T\omega \tau _w}e^{-\zeta a_T}
\sum_{k=0}^{\infty} \left( \frac{\zeta b_T}4\right) ^k\frac{p_k\left( \alpha _T\omega
\tau _w\right) }{\left( k!\right) ^2}  \label{e92}
\end{equation}
where $p_k$ is a polynomial function of frequency:

\begin{equation}
p_k\left( x\right) =\sum_{j=0}^{k}\frac{\left(
2x\right) ^j\left( 2k-j\right) !}{j!\left( k-j\right) !}  \label{93}
\end{equation}
This leads to a normalized spectrum:

\begin{equation}
\bar{S}_{NN}\left( T,\omega \right) =\pi \alpha _T\tau
_we^{-\alpha _T\omega \tau _w}e^{-\frac{\zeta b_T}{\alpha _T^2}}
\sum_{k=0}^{\infty}\left( \frac{\zeta b_T}4\right) ^k\frac{%
p_k\left( \alpha _T\omega \tau _w\right) }{\left( k!\right) ^2}  \label{94}
\end{equation}
with a peak value at zero frequency:

\begin{equation}
\bar{S}_{NN}\left( T,0\right) =\pi \alpha _T\tau _we^{-\frac{%
\zeta b_T}{\alpha _T^2}}\sum_{k=0}^{\infty} \left( 
\frac{\zeta b_T}4\right) ^k\frac{\left( 2k\right) !}{\left( k!\right) ^3}
\label{e95}
\end{equation}

The expressions (\ref{e92}-\ref{e95}) provide a quantitative evaluation of
the fluctuations of the number of atoms coupled to the probe beam in the
general case of a motion under gravity. They may be used for assessing the
effect of these number fluctuations in any high precision measurements such
as spectroscopic, nonlinear or quantum optics experiments where the falling
cloud acts as the atomic medium. For only qualitative assessments, we can
draw the following two conclusions: first, the instantaneous variance of the
effective atomic number has a subPoissonian character; it is precisely half
the Poissonian variance at the limit of a small probe beam size. Second, the
correlation time for the fluctuations of the effective atomic number is
mainly determined by the time of flight $\tau _w$ through the probe beam.

\section{\bf{Fluctuations of the cavity detuning}}

We now focus our attention on experiments where the probe field goes through
an optical cavity containing the cloud of cold atoms.\ As already discussed
in the introduction, atomic number fluctuations emulate fluctuations of the
cavity detuning.\ The present section is devoted to a precise evaluation of
this effect.\ For simplicity, we consider only the regime of a linear
polarisability in the limit where the Rayleigh divergence length is much
larger than the cloud size (same simplifying assumptions as in section 5).\
We also consider the dispersive regime where the atom is excited far from
resonance ($\delta ^2\gg 1$).

A key parameter in the experiments with atoms inside a cavity is the
so-called cooperativity parameter (see for example \cite{Bistab1,Hilico} and
references therein):

\begin{equation}
C=\frac{3\lambda ^2}{4\pi S}\frac n{2\kappa \tau _c}  \label{e96}
\end{equation}
where $\kappa $ is the decay rate of the intracavity field and $\tau _c$ the
cavity round-trip time ($2\kappa \tau _c$ is just the intensity transmission
of the coupling mirror), while $S$ and $n$ are respectively the beam section
and the effective atomic number studied in the present paper.\ The effect of
the atomic medium on the field may be written as a variation $\Phi $ of the
cavity detuning (see for example \cite{Hilico}):

\begin{equation}
\Phi =\frac{2\kappa C}\delta =\frac{3\lambda ^2}{4\pi S}\frac n{\delta \tau
_c}  \label{97}
\end{equation}

The fluctuations of the atomic number $N$ studied in the previous section
are therefore equivalent to fluctuations of the cavity detuning $\Phi $
characterized by a noise spectrum:

\begin{equation}
S_{\Phi \Phi }\left( T,\omega \right) =\left( \frac{3\lambda ^2}{4\pi S}%
\right) ^2\frac{S_{NN}\left( T,\omega \right) }{\left( \delta \tau _c\right)
^2}  \label{98}
\end{equation}
that is (using equations \ref{e88} and \ref{e96}):

\begin{equation}
S_{\Phi \Phi }\left( T,\omega \right) =\kappa \frac{C\left( T\right) }{%
\delta ^2}\frac{3\lambda ^2}{4\pi S}\frac{\bar{S}_{NN}\left(
T,\omega \right) }{\tau _c}  \label{e99}
\end{equation}
We have introduced here a time-dependent cooperativity parameter $C(T)$
proportional to the mean effective number $n(T)$ at time $T$.\ Since the
cavity detuning $\Phi $ is a frequency, its noise spectrum $S_{\Phi \Phi }$
also has the dimension of a frequency.\ The first factor appearing in its
expression (\ref{e99}) is the frequency $\kappa $ , which can be regarded as
measuring the noise spectrum of detuning fluctuations associated with the
decay of the intracavity field through the coupling mirror.\ The other
factors which appear in equation (\ref{e99}) have been written in
dimensionless terms.

As long as $S_{\Phi \Phi }$ remains smaller than $\kappa $, the effect of
the detuning fluctuations on cavity dynamics remains in a linear regime.\ In
the opposite case, it would be necessary to give a more elaborate treatment
of these fluctuations accounting for multiplicative noise processing.\ It is
worth stressing that detuning fluctuations may not be considered as a white
noise, since the spectrum has a width essentially given by the inverse of
the time of flight through the detection beam.

In the case where the measurements are performed at frequencies much higher
than this spectral width, the detuning fluctuations become negligible. This
is in particular the case in the experiment on squeezing with cold atoms 
\cite{Squeezing94,Squeezing95} where the squeezing spectrum is monitored at
frequencies in the MHz range. It follows that the effective atomic number
may be considered as a fluctuationless classical variable, even when it
varies as a function of time while the cloud is exploding and falling down.
The fluctuations of the effective number might be detectable at much lower
frequencies in the kHz range. In order to detect these fluctuations, it
would thus be necessary to get rid of the large excess noise present in this
frequency range.

In conclusion, we have investigated the problem of evaluating the number of
atoms effectively interacting with a probe beam passing through a cloud of
cold atoms released from a magneto-optical trap, accounting for the
transverse Gaussian profile of the probe beam, in the general case of a
motion under gravity.

We have shown that this problem requires a detailed treatment which depends
on the quantity of interest in the specific measurement. We have given
explicit evaluations of the effective atomic number for the two cases where
the linear or nonlinear phaseshifts are measured. These two cases correspond
to a same time variation of the effective number when the two simplifying
assumptions of a small beam waist ($w\ll \sigma _r$) and of a long Rayleigh
divergence length ($\sigma _r\ll l_R$) are fulfilled. The more general
situation may also be dealt with by using the expressions obtained in
sections 3 and 4.

We have also calculated the correlation functions characterizing the
fluctuations of the effective atomic number, focussing attention upon
fluctuations of the linear phaseshift since they may be quite large and
consequently affect high precision measurements such as spectroscopic,
nonlinear or quantum optics experiments. We have found that the
instantaneous variance of the effective atomic number has a subPoissonian
character, due to the Gaussian profile of the probe beam. In particular, the
variance is half the Poissonian variance at the limit of a small beam waist (%
$w\ll \sigma _r$). At the same limit, the spectral width characteristic of
these fluctuations is essentially the inverse of the time of flight through
the probe beam for an atom flying at the mean thermal velocity. The effect
of these number fluctuations on the field fluctuations is thus confined to a
low frequency range, especially for cold atoms which correspond to
relatively long time of flight. For atoms at room temperature, the spectral
width would be larger. In both cases, the techniques used in the present
paper allow to obtain quantitative estimates of the influence of atomic
number fluctuations on field fluctuations, and then to the sensitivity of
optical measurements.

{\bf{Acknowledgements}}

Thanks are due to Jean Dalibard and Claude Fabre for discussions.

\end{document}